\documentclass[11pt]{article}
\usepackage[DIV13]{typearea}
\usepackage{amsmath} 
\usepackage{amssymb}   
\usepackage{amsfonts}
\usepackage{amscd}
\usepackage{graphicx}
\usepackage[footnotesize,bf]{caption2}
\usepackage{cite}
\usepackage{bbold}
\usepackage{longtable}
\usepackage[center,footnotesize,hang]{subfigure}
\usepackage{url}
\usepackage{epsfig}
\usepackage{dcolumn}
\usepackage{bm}

\newcommand{\eV}{\ensuremath{\:\mathrm{eV}}}

\newcommand{\GeV}{\ensuremath{\:\mathrm{GeV}}}

\newcommand{\Delmsol}{\Delta \mathrm{m}_{21} ^{2}}
\newcommand{\Delmatm}{|\Delta \mathrm{m}_{31} ^{2}|}

\newcommand{\Eqref}[1]{Eq.\eqref{#1}}

\newcommand{\Tabref}[1]{Table \ref{#1}}

\newcommand{\Secref}[1]{Section \ref{#1}}
\newcommand{\Appref}[1]{Appendix \ref{#1}}

\newcommand{\MoreRep}[2]{\underline{\mbox{\textbf{#1}}} _{\mbox{\textbf{#2}}}}
\newcommand{\Groupname}[2]{$ {#1} _{#2} $}

\newcommand{\VEV}[1]{\langle #1 \rangle}
\newcommand{\La}{\Lambda}

\begin{document}

\begin{titlepage}
\begin{flushright}
DFPD-2011/TH/12\\
RM3-TH/11-8 
\end{flushright}
\vspace*{5mm}

\begin{center}
{\Large\sffamily\bfseries
\mathversion{bold} $D_{14}$ - A Common Origin of the Cabibbo Angle and\\ the Lepton Mixing Angle $\theta_{13}^l$
\mathversion{normal}} \\[13mm]
{\large C. Hagedorn$^a$ and D. Meloni$^b$
\\[5mm]
{\small \textit{$^a$Dipartimento di Fisica e Astronomia "G. Galilei", Universit\`a di Padova\\
INFN, Sezione di Padova, Via Marzolo 8, I-35131 Padua, Italy\\
$^b$Dipartimento di Fisica "E. Amaldi", Universit\`a degli Studi Roma Tre, \\ Via della Vasca Navale 84, 00146 Roma, Italy
}}}
\vspace*{1.0cm}
\end{center}
\normalsize
\begin{abstract}
\noindent 
It has been shown that the Cabibbo angle can be predicted  in terms of group theoretical quantities, if the dihedral group $D_{14}$
plays the role of a flavor symmetry. We 
extend a supersymmetric $D_{14}$ model to the lepton sector and show that  $\theta_{13}^\nu$ and the deviation of $\theta_{23}^\nu$ from maximal mixing in the neutrino sector originate, 
similar to the Cabibbo angle in the quark sector, from a mismatch of different subgroups of $D_{14}$ and are of the size of the Cabibbo angle. The mixing angles in the charged lepton sector are small.
Thus, the lepton mixing angle $\theta_{13}^l$ is naturally in its experimentally preferred range
and $\theta_{23}^l$ within its $3 \sigma$ range. The solar mixing angle is of order one and 
the charged lepton mass hierarchy is correctly reproduced. 
Leading order results are only slightly perturbed, if next-to-leading order corrections are taken into account.
\end{abstract}
\end{titlepage}
\setcounter{footnote}{0}

\section{Introduction}
\label{sec:intro}

The properties of fermion masses and mixing are very puzzling and cannot be predicted within the Standard Model (SM). 
The peculiar mixing pattern observed in the lepton sector with two large mixing angles and one small one is in sharp contrast to the one
of the quark sector where only the Cabibbo angle $\theta_C \approx \lambda \approx 0.22$ is non-negligible. The charged fermion masses are 
strongly hierarchical, while neutrino masses have a much milder hierarchy (and might have an inverted ordering). Symmetries relating 
the different generations of fermions are known to be a very useful tool for explaining (some of) these peculiar features. 
In particular, finite discrete groups which are broken in a non-trivial way are capable of predicting
fermion mixing, see e.g. \cite{thetaC_lam,dntheory,dATFH2}. Among the many possible symmetries to choose from, see for reviews \cite{reviews}, we concentrate on a dihedral group.
 As has been shown, the value of the Cabibbo angle $\theta_C$
can be explained with the flavor groups $D_7$ \cite{thetaC_lam,dntheory},\footnote{The group $D_7$ has also been used as flavor symmetry in \cite{anotherD7}.}
 $D_{14}$ \cite{D14_quarks} or with the group $D_{12}$  \cite{Kim:2010zub}. 
Similarly, it has been shown that $\mu\tau$ symmetric lepton mixing,
$\theta_{23}^l=\pi/4$ and $\theta_{13}^l=0$, can  originate from the dihedral groups $D_3$ \cite{D3mutau} and $D_4$ \cite{D4mutau}, while the golden ratio mixing pattern
with $\theta_{12}^l=\pi/5$ can arise from the dihedral group $D_{10}$ \cite{Adulpravitchai:2009bg}.
A crucial ingredient is the breaking of the flavor group to distinct subgroups in different sectors of the theory.
This mismatch is the source of quark and lepton mixing.

We present a model with the flavor group $D_{14}$ in which the Cabibbo angle as well as $\theta_{13}^l$ of the order of the Cabibbo angle and
$\theta_{23}^l$ within its experimentally allowed $3 \sigma$ range are predicted through a particular breaking of $D_{14}$. 
As framework we use the Minimal Supersymmetric SM (MSSM). We add gauge singlets, so-called
flavons, which only transform under the flavor group, to break the latter correctly. Left-handed quarks and leptons as well as right-handed neutrinos
are assigned to a singlet and a doublet representation of $D_{14}$. The crucial difference is that the first two generations of left-handed quarks form a doublet under $D_{14}$, 
while in the case of left-handed leptons and right-handed neutrinos the second and third generations are unified into one doublet of $D_{14}$.
 In order to properly segregate the different $D_{14}$ breaking sectors we employ a $Z_7$ symmetry which does not distinguish between different generations. 
 At leading order (LO), we predict the Cabibbo angle $\theta_C$ to fulfill $\sin \theta_C \approx \sin \pi/14 \approx 0.22$ through the breaking of $D_{14}$
 to different $Z_2$ subgroups in the down and up quark sectors. In the neutrino sector, 
this mismatch is the origin of $\theta_{13}^\nu \approx \mathcal{O}(\lambda)$ and $\theta_{23}^\nu- \pi/4 \approx  \mathcal{O}(\lambda)$, because
the right-handed neutrino mass matrix is governed by the $Z_2$ symmetry preserved in the down quark sector, whereas the flavons associated with the up quark sector, whose 
vacuum expectation values (VEVs) conserve a different $Z_2$ symmetry, are dominantly responsible for the structure of the Dirac neutrino mass matrix. 
The solar mixing angle is generically of order one.
The angles $\theta_{13}^q$ and $\theta_{23}^q$ in the quark sector as well as the mixing angles in the charged lepton sector are small, of the order of the generic 
expansion parameter $\epsilon \approx 0.04$. As a consequence, the atmospheric mixing angle $\theta_{23}^l$ is within its $3 \sigma$ range and $\theta_{13}^l \approx \mathcal{O}(\lambda)$,
in accordance with the latest experimental findings \cite{T2K,MINOS,DC,DayaBay,RENO} and global fit results \cite{fogli,maltoni,schwetz}.
In order to fully explain the mass hierarchy among quarks we invoke a Froggatt-Nielsen (FN) symmetry \cite{FN}, while the hierarchy among the charged lepton
masses is derived with the help of discrete symmetries only. 
The generation of the mass hierarchies is facilitated by assigning the right-handed charged fields to singlets under $D_{14}$.
The light neutrino mass spectrum can have either hierarchy in our model.
Next-to-leading order (NLO) corrections, arising from various operators with several flavons, are shown to
affect the LO results only slightly. 
 
The paper is structured as follows: in \Secref{sec:outline} we outline the setup of our model and show its
particle content. \Secref{sec:leptons} contains the discussion of the lepton sector at LO and NLO, showing how $\theta_{13}^\nu \approx \mathcal{O}(\lambda)$
and $\theta_{23}^\nu - \pi/4 \approx  \mathcal{O}(\lambda)$ originate.
The results for the quark sector, which essentially coincide with those of the model \cite{D14_quarks}, are briefly discussed in \Secref{sec:quarks}. The flavon superpotential
and the vacuum alignment can be found in \Secref{sec:flavons}. We discuss the relevant subgroups of $D_{14}$ in some detail in \Secref{sec:D14sub}.
We summarize our results in \Secref{sec:summary}. The basics of the group theory of $D_{14}$ are given in \Appref{app:grouptheory}.

\section{Outline of the model}
\label{sec:outline}

The flavor symmetry of the model is the direct product of the dihedral group $D_{14}$, an FN symmetry $U(1)_{FN}$ and the cyclic group $Z_7$.
The assignment of quarks, flavons responsible for quark masses, the MSSM Higgs doublets $h_{u,d}$ 
and the FN field $\theta$ is mainly adopted from \cite{D14_quarks}. In comparison to \cite{D14_quarks}, we extend the cyclic symmetry from $Z_3$ to $Z_7$ and define 
 $\omega_7=\mathrm{e}^{\frac{2 \pi i}{7}}$. Quark $SU(2)_L$ doublets, the MSSM Higgs doublet $h_d$, flavons belonging to the set
$\{ \psi^{u}_{1,2}, \chi_{1,2}^{u}, \xi^{u}_{1,2}, \eta^{u}\}$ and the FN field $\theta$
are uncharged under $Z_7$, while the right-handed down quarks acquire a phase $\omega_7$
and flavons giving mainly masses to down quarks a phase $\omega_7^6$. In contrast to the original proposal \cite{D14_quarks},
the MSSM Higgs $h_u$ and the right-handed quarks $u^c$, $c^c$, $t^c$ carry in the present setup the $Z_7$ charge 
$\omega_7^3$ and $\omega_7^4$, respectively. This is summarized in  \Tabref{tab:quarks}.
\begin{table}
\hspace{-0.8in}
\parbox{6in}{
\begin{tabular}{|c||c|c|c|c|c|c|c|c||c|c||c|c|c|c||c|c|c|c|c|}\hline
Field & $Q_D $ & $Q_{3}$ & $u^c$ & $c^c$ & $t^c$ & $d^c$ & $s^c$ & 
$b^c$ & $h_{u}$ & $h_{d}$ & 
$\psi^{u}_{1,2}$ & $\chi_{1,2}^{u}$ & $\xi^{u}_{1,2}$ & $\eta^{u}$ &
$\psi^{d}_{1,2}$ & $\chi_{1,2}^{d}$ & $\xi^{d}_{1,2}$ & $\eta^{d}$ & $\sigma$\\ 
\hline
\Groupname{D}{14} & $\MoreRep{2}{1}$ & $\MoreRep{1}{1}$ & $\MoreRep{1}{4}$ & 
$\MoreRep{1}{3}$ & $\MoreRep{1}{1}$  & $\MoreRep{1}{3}$ & 
$\MoreRep{1}{1}$ & $\MoreRep{1}{4}$ & $\MoreRep{1}{1}$ & $\MoreRep{1}{1}$ & $\MoreRep{2}{1}$ & $\MoreRep{2}{2}$ &
$\MoreRep{2}{4}$ & $\MoreRep{1}{3}$ & $\MoreRep{2}{1}$ & $\MoreRep{2}{2}$ &
$\MoreRep{2}{4}$ & $\MoreRep{1}{4}$ & $\MoreRep{1}{1}$\\
\Groupname{Z}{7} & $1$ & $1$ & $\omega_7^4$ & $\omega_7^4$ & $\omega_7^4$ & $\omega_7$ & $\omega_7$ & $\omega_7$
& $\omega_7^3$ & $1$ & $1$ & $1$ & $1$ & $1$ & $\omega_7^6$ & $\omega_7^6$ & $\omega_7^6$ & $\omega_7^6$ & $\omega_7^6$\\
$U(1)_{FN}$ & 0 & 0 & 2 & 0 & 0 & 1 & 1 & 0 & 0 & 0 & 0 & 0 & 0 & 0 & 0 & 0 & 0 & 0 & 0\\ 
\hline
\end{tabular}}
\begin{center}
\normalsize
\begin{minipage}[t]{15cm}
\caption[Quarks of the Model]{\it Transformation properties of the fields associated with the quark sector under $D_{14} \times Z_7 \times U(1)_{FN}$. 
The left-handed quark
doublets are denoted by $Q_D = (Q_{1},Q_{2})^T$, $Q_{1}=(u,d)^{T}$, $Q_{2}=(c,s)^{T}$, $Q_{3}=(t,b)^{T}$ 
and the right-handed quarks by $u^c$, $c^c$, $t^c$
and $d^c$, $s^c$, $b^c$. The flavon fields indexed by a $u$ give masses to the up quarks, at lowest order. 
Similarly, the fields which carry an index $d$ (including the field $\sigma$)
couple to down quarks at this order.
A field $\theta$ being a gauge singlet and
transforming trivially under $D_{14} \times Z_7$
is responsible for the 
breaking of the $U(1)_{FN}$ symmetry. Its charge under
$U(1)_{FN}$ is taken to be $-1$. 
$\omega_7$ is the seventh root of unity $\omega_7=\mathrm{e}^{\frac{2 \pi i}{7}}$.
\label{tab:quarks}}
\end{minipage}
\end{center}
\end{table}
 Similar to the left-handed quarks, the lepton $SU(2)_L$ doublets $L$ and the right-handed neutrinos $\nu^c$
 are assigned to a one- and a two-dimensional representation of $D_{14}$. In order to motivate the largeness of the atmospheric 
 mixing angle (which is the largest mixing angle in the lepton sector), however, we unify the second and third generations into
a doublet, $L_D \sim \MoreRep{2}{2}$ and $\nu^c_D \sim \MoreRep{2}{1}$, respectively, instead of the first two ones.
The three generations of right-handed charged leptons are assigned to the trivial singlet $\MoreRep{1}{1}$. 
The Dirac neutrino mass requires the insertion of one flavon belonging to the set
$\{\psi^{u}_{1,2}, \chi_{1,2}^{u}, \xi^{u}_{1,2}, \eta^{u} \}$, because left-handed leptons and right-handed neutrinos are 
in different $D_{14}$ representations. Majorana masses of the 
right-handed neutrinos stem at LO from couplings to the down-type flavons $\{\psi^{d}_{1,2}, \chi_{1,2}^{d}, \xi^{d}_{1,2}, \eta^{d}, \sigma \}$, 
since right-handed neutrinos have charge $\omega_7^4$ under
$Z_7$. The resulting light neutrino mass matrix has a non-hierarchical structure which
is compatible with either mass hierarchy.
Left-handed leptons are uncharged under $Z_7$, while
right-handed charged leptons transform as $\omega_7^5$. Due to this and due to the $D_{14}$ assignment of the fields, the operators, giving masses to charged leptons, 
have to contain at least one flavon. For this purpose, we introduce 
the fields $\chi^e_{1,2}$, which form the doublet $\MoreRep{2}{2}$ under $D_{14}$ and acquire a phase  
$\omega_7^2$ under $Z_7$. The tau lepton mass then originates, like the mass of the bottom quark, from non-renormalizable terms.
As a consequence, small and moderate values of $\tan \beta$, $\tan \beta = \langle h_u \rangle/\langle h_d \rangle$, are preferred.
The muon mass is generated
through the insertion of two flavons, $\chi^e_{1,2}$ and one of the up-type flavons, while the electron
mass only arises at the level of three flavon insertions. Shifts in the VEVs of the flavons
also contribute to the muon mass at the same level as operators with two flavons.
The hierarchy among the charged leptons is, hence, reproduced without
invoking the FN symmetry. 
All non-renormalizable terms are suppressed by (powers of) the cutoff scale $\Lambda$ which is expected to be 
much larger than the electroweak scale and is related to the light neutrino mass scale, see \Eqref{eq:Lambda_estimate}, in the present model.  
The transformation properties under $D_{14} \times Z_7$
of the lepton fields and $\chi^e_{1,2}$ are collected in \Tabref{tab:leptons}.
\begin{table}
\begin{center}
\begin{tabular}{|c||c|c|c|c|c|c|c||c|}\hline
Field & $L_1 $ & $L_D$ & $e^c$ & $\mu^c$ & $\tau^c$ & $\nu^c_1$ & $\nu_D^c$ & $\chi^{e}_{1,2}$\\ 
\hline
\Groupname{D}{14} & $\MoreRep{1}{3}$ & $\MoreRep{2}{2}$ & $\MoreRep{1}{1}$ & 
$\MoreRep{1}{1}$ & $\MoreRep{1}{1}$  & $\MoreRep{1}{1}$ & $\MoreRep{2}{1}$ & $\MoreRep{2}{2}$ \\
 \Groupname{Z}{7} & $1$ & $1$ & $\omega_7^5$ & $\omega_7^5$ & $\omega_7^5$ & $\omega_7^4$ & $\omega_7^4$ & $\omega_7^2$\\
\hline
\end{tabular}
\end{center}
\begin{center}
\normalsize
\begin{minipage}[t]{15cm}
\caption[Leptons of the Model]{\it Fields associated with the lepton sector of the model.
The lepton $SU(2)_L$ doublets are called $L_1 = (\nu_e, e)^{T}$ and $L_D= (L_2, L_3)^{T}$
with $L_2 = (\nu_\mu, \mu)^{T}$ and $L_3 = (\nu_\tau, \tau)^{T}$. The right-handed 
fields are denoted by $e^c$, $\mu^c$, $\tau^c$ for the charged leptons and 
$\nu^c_1$ and $\nu^c_D = (\nu^c_2, \nu^c_3)^{T}$ are right-handed neutrinos.
Only one further flavon multiplet $\chi^e_{1,2}$ is added which mainly gives masses to charged leptons.
All fields contained in this table are neutral with respect to the FN symmetry.
\label{tab:leptons}}
\end{minipage}
\end{center}
\end{table}

We can write down the superpotential $w$ which consists of three parts
\begin{equation}
w = w_q + w_l + w_f \; .
\end{equation}
$w_q$ ($w_l$) contains the Yukawa couplings of quarks (leptons) and $w_f$ the superpotential responsible
for the vacuum alignment of the flavons.  Here, we only list the LO vacuum structure,
 necessary for the computation of the main contributions to the fermion mass matrices and postpone any further discussion of the vacuum alignment to \Secref{sec:flavons}.
The LO VEVs of the fields 
$\psi^{u}_{1,2}$, $\chi^{u}_{1,2}$ and $\xi^u_{1,2}$, which preserve a $Z_2$ symmetry (called $Z_2^u$ in the following) generated by the element $\mathrm{B}$ of $D_{14}$, 
for details see Sections \ref{sec:flavons}, \ref{sec:D14sub} 
and \Appref{app:grouptheory}, are of the form
\begin{equation}
\label{eq:vacuumup_LO}
\!\!\!\!\!\!\! \!\!\!\!\left( \begin{array}{c} \VEV{\psi_1^u} \\ \VEV{\psi_2^u} \end{array} \right) = v^u \left( \begin{array}{c} 
1 \\ 1 \end{array} \right) , \;
\left( \begin{array}{c} \VEV{\chi_1^u} \\ \VEV{\chi_2^u} \end{array} \right) = w^u \, 
\left( \begin{array}{c} 1 \\  1 \end{array} \right) , \;
\left( \begin{array}{c} \VEV{\xi_1^u} \\ \VEV{\xi_2^u} \end{array} \right) = z^u \,  
\left( \begin{array}{c} 1\\  1 \end{array} \right)
\end{equation}
together with $\langle \eta^u \rangle \neq 0$.  Being correlated through the superpotential, see \Secref{sec:flavons} and \cite{D14_quarks}, we expect a common order of magnitude of the VEVs:
 $v^u \sim w^u \sim z^u \sim \langle \eta^u \rangle$.
The down-type flavons, whose VEVs preserve at LO a $Z_2$ subgroup (called $Z_2^d$ in the following) generated by $\mathrm{B} \mathrm{A}^{k}$ with $k=1,3,\dots,13$,
are of the form
\begin{equation}
\label{eq:vacuumdown_LO} 
\!\!\!\!\!\!\!\!\! 
\left( \begin{array}{c} \VEV{\psi_1^d} \\ \VEV{\psi_2^d} \end{array} \right) = v^d \left( \begin{array}{c} \mathrm{e}^{-2 i \gamma k} \\ 1 \end{array} \right) \; , \;\;
\left( \begin{array}{c} \VEV{\chi_1^d} \\ \VEV{\chi_2^d} \end{array} \right) = w^d \mathrm{e}^{2 i \gamma k} \left( \begin{array}{c} \mathrm{e}^{-4 i \gamma k} \\ 1 \end{array} \right) \; , \;\;
\left( \begin{array}{c} \VEV{\xi_1^d} \\ \VEV{\xi_2^d} \end{array} \right) = z^d \mathrm{e}^{4 i \gamma k} \left( \begin{array}{c} \mathrm{e}^{-8 i \gamma k} \\ 1 \end{array} \right) 
\end{equation}
with $\VEV{\eta^d}$ and $\VEV{\sigma}$ being non-zero and
\begin{equation}
\gamma = \frac{\pi}{14} \, .
\end{equation} 
 Also these VEVs are correlated through the parameters of the superpotential and thus are of the
same order of magnitude: $v^d \sim w^d \sim z^d \sim \langle \eta^d \rangle \sim \langle \sigma \rangle$. The parameter $k$ has to be chosen as $k=1$ or $k=13$
 for reproducing correctly the Cabibbo angle \cite{D14_quarks}. In the following we take $k=1$. 
The VEV of the fields $\chi^e_{1,2}$ reads at LO
\begin{equation}
\label{eq:align_che}
\left( \begin{array}{c} \VEV{\chi_1^e} \\ \VEV{\chi_2^e}\end{array} \right)
= v^e \left( \begin{array}{c} 1 \\ 0 \end{array} \right)
\;\;\; \mathrm{or} \;\;\;
\left( \begin{array}{c} \VEV{\chi_1^e} \\ \VEV{\chi_2^e}\end{array} \right)
= v^e \left( \begin{array}{c} 0 \\ 1 \end{array} \right) \; .
\end{equation}
These vacua both break $D_{14}$ to the subgroup $Z_2$ generated by the element $\mathrm{A}^7$ (which is obviously distinct from the groups $Z_2^u$ and $Z_2^d$), 
because $\chi^e_{1,2}$ transform as an unfaithful representation of $D_{14}$, see
\Secref{sec:D14sub}. As can be checked, the results of fermion masses and mixing are independent of the actual choice of the vacuum structure of $\chi^e_{1,2}$ and without loss of 
generality we can assume the first one in \Eqref{eq:align_che} to be realized.
We take all flavon VEVs to be of the same order of magnitude $\epsilon \, \Lambda$.  
The expansion parameter $\epsilon$ is determined to be around $0.04$ in order to correctly reproduce the charged fermion mass hierarchy.
The VEV of the FN field $\theta$ is also taken to be of order $\epsilon \, \Lambda$: $\langle \theta \rangle \equiv t \, \Lambda \sim \epsilon \, \Lambda$.

\section{Lepton sector}
\label{sec:leptons}

In this section we present the leading and subleading results for the charged lepton and the neutrino sectors and show
that $\theta_{13}^l \approx \mathcal{O}(\lambda)$ and $\theta_{23}^l -\pi/4 \approx  \mathcal{O}(\lambda)$
are achieved. This result can be traced back in our model to the presence of the remnant $Z_2^d$ subgroup, the one
preserved in the down quark sector, in the right-handed neutrino mass matrix, whereas the Dirac neutrino mass
matrix is determined by the $Z_2^u$ symmetry, conserved in the up quark sector.\footnote{Here and in the following we focus
on the subgroups of $D_{14}$ and the invariance of the mass matrices under the latter and do not discuss
accidental symmetries which are possibly present - especially when considering the mass matrices only at LO.} 
The solar mixing angle
is generically large, but not fixed by the properties of the group $D_{14}$. Subleading corrections arising from NLO terms in the neutrino sector
as well as from the small mixing in the charged lepton sector (at maximum, of order $\epsilon$) only slightly influence these results.

\subsection{Charged leptons}
\label{subsec:chargedleptons}

The lowest order operators in the charged lepton sector are 
\begin{equation}
\label{eq:chleptons_LO}
\frac{1}{\La} (L_D \chi^e) e^c h_d + \frac{1}{\La} (L_D \chi^e) \mu^c h_d + \frac{1}{\La} (L_D \chi^e) \tau^c h_d \; .
\end{equation}
Here and in the following we omit order one coefficients in front of the operators and $\left( \dots \right)$ denotes the contraction
to the trivial singlet of $D_{14}$. 
The operators in \Eqref{eq:chleptons_LO} only give rise to the elements of the third row of the charged lepton mass matrix $\mathcal{M}_{e}$
(in the left-right basis), if we use as alignment of $\langle \chi^e_{1,2} \rangle$ the first one found in \Eqref{eq:align_che}. Thus, we generate only
the tau lepton mass. It is of order $\epsilon \, \langle h_d \rangle$ and is correctly predicted for small and moderate values of $\tan\beta$.
The mass of the muon arises from the shift in the vacuum of the fields
$\chi^e_{1,2}$ as well as from two flavon insertions involving $\chi^e_{1,2}$ and one up-type flavon:
\begin{eqnarray}\nonumber
&& \frac{1}{\La} (L_D \delta \chi^e) e^c h_d + \frac{1}{\La} (L_D \delta \chi^e) \mu^c h_d + \frac{1}{\La} (L_D \delta \chi^e) \tau^c h_d \\ 
&+& \frac{1}{\La^2} (L_D \chi^e \xi^u) e^c h_d + \frac{1}{\La^2} (L_D \chi^e \xi^u) \mu^c h_d 
+ \frac{1}{\La^2} (L_D \chi^e \xi^u) \tau^c h_d 
\end{eqnarray}
with $\delta \chi^e$ indicating the insertion of the shifted vacuum of the fields $\chi^e_{1,2}$.
Under the assumption that the VEV shift of $\chi^e_{1,2}$ is of the order $\epsilon \, v^e \approx \epsilon^2 \La$, see \Secref{sec:flavons},
these operators lead to entries of the second row of $\mathcal{M}_e$ proportional to $\epsilon^2$. Thus, the ratio of muon to tau lepton mass,
$m_\mu/m_\tau \sim \mathcal{O}(\epsilon)$, is correctly reproduced. The electron mass is only generated, if insertions
of three flavons are considered, two of them being up-type flavons and the third one necessarily being $\chi^e_{1,2}$. The operators
generating the elements of the first row of $\mathcal{M}_e$ read 
\begin{eqnarray}\nonumber
&& \frac{1}{\La^3} (L_1 \chi^e \psi^u \xi^u) e^c h_d + \frac{1}{\La^3} (L_1 \chi^e \psi^u \xi^u) \mu^c h_d 
+ \frac{1}{\La^3} (L_1 \chi^e \psi^u \xi^u) \tau^c h_d \\ 
&+& \frac{1}{\La^3} (L_1 \eta^u) (\chi^e \chi^u) e^c h_d + \frac{1}{\La^3} (L_1 \eta^u) (\chi^e \chi^u) \mu^c h_d
+  \frac{1}{\La^3} (L_1 \eta^u) (\chi^e \chi^u) \tau^c h_d \, .
\end{eqnarray}
Thus, the correct mass ratio $m_e/m_\tau \sim {\cal O}(\epsilon^2)$ is achieved.
Further operators arising at the three flavon level contributing to the elements of the second and third rows of $\mathcal{M}_{e}$ 
are subleading. 
Eventually, the most general structure of the charged lepton mass matrix in our model is 
\begin{equation}
\label{eq:Me}
\mathcal{M}_e = \left( \begin{array}{ccc}
		\alpha^e_1 \, \epsilon^3 & \alpha^e_2 \, \epsilon^3 & \alpha^e_3 \, \epsilon^3\\
		\alpha^e_4 \, \epsilon^2 & \alpha^e_5 \, \epsilon^2 & \alpha^e_6 \, \epsilon^2\\
		\alpha^e_7 \, \epsilon & \alpha^e_8 \, \epsilon & \alpha^e_9 \, \epsilon
\end{array}
\right) \langle h_d \rangle \, ,
\end{equation}
where all coefficients $\alpha^e_i$ are in general complex with an absolute value of order one.
As can be read off from \Eqref{eq:Me}, the mixing angles of the left-handed charged leptons are small 
\begin{equation}
\label{chmix}
 \theta_{12}^e \sim \mathcal{O}(\epsilon)  \qquad , \qquad \theta_{13}^e   \sim \mathcal{O}(\epsilon^2) \qquad , \qquad \theta_{23}^e \sim \mathcal{O}(\epsilon) \, .
\end{equation}
In contrast to this, the (unobservable) mixing of the right-handed charged leptons is sizable due to the lopsided structure of the mass matrix $\mathcal{M}_e$
\cite{lopsided}.

\subsection{Neutrinos}
\label{subsec:neutrinos}

In the following, we discuss the structure of the Majorana mass matrix of the right-handed neutrinos, of the Dirac neutrino mass matrix and the
contribution of the Weinberg operator to the light neutrino mass matrix at LO and NLO. We explicitly show that $\theta_{13}^\nu$ and $\theta_{23}^\nu$
deviate by ${\cal O}(\lambda)$ from the $\mu\tau$ symmetric result, while the angle $\theta_{12}^\nu$ is generically of order one.

\subsubsection{LO results}

At LO, i.e. at the one flavon level, the operators
\begin{equation}
\label{MRinv}
\nu^c_1 \nu^c_1 \sigma + (\nu^c_D \nu^c_D) \sigma 
\end{equation}
and
\begin{equation}
\label{MRtheta13}
\nu^c_1 (\nu^c_D \psi^d) + (\nu^c_D \nu^c_D \chi^d)
\end{equation}
contribute to the right-handed neutrino mass matrix ${\cal M}_R$. If one plugs in the LO flavon VEVs, one sees that 
the contributions coming from the terms in Eq.(\ref{MRinv}) cannot break the group $D_{14}$, while
those arising from the terms in Eq.(\ref{MRtheta13}) preserve only the symmetry $Z_2^d$.
The matrix ${\cal M}_R$ takes the form 
\footnote{We have defined the coupling $\alpha^M_3$ in such a way that the phase $\mathrm{e}^{\pm i \gamma}$ appears symmetrically
in the (12) and (13) elements of the matrix ${\cal M}_R$.}
\begin{equation}
\mathcal{M}_R 
= \left( \begin{array}{ccc}
 \alpha^M_1 & \alpha^M_3  \, \mathrm{e}^{i \gamma}\ & \alpha^M_3 \, \mathrm{e}^{-i \gamma}\\
  \alpha^M_3 \,  \, \mathrm{e}^{i \gamma} &  \alpha^M_4 \,  \, \mathrm{e}^{2 i \gamma}\ & \alpha^M_2\\
  \alpha^M_3  \, \mathrm{e}^{-i \gamma}\ & \alpha^M_2 &  \alpha^M_4 \,  \, \mathrm{e}^{-2 i \gamma}\
\end{array}
\right) \, \epsilon \, \Lambda  \, .
\end{equation}
The Dirac neutrino mass matrix $\mathcal{M}^D_\nu$ is also dominantly generated through one flavon insertions
\begin{equation}
\label{Diracnu_LO}
\frac{1}{\La} (L_1 \eta^u) \nu^c_1 h_u + \frac{1}{\La} (L_D \chi^u) \nu^c_1 h_u
+ \frac{1}{\La} (L_D \nu^c_D \psi^u) h_u 
\end{equation}
and it is of the form
\begin{equation}
\mathcal{M}^D_\nu = \left( \begin{array}{ccc}
  \alpha^D_1 & 0 & 0\\
\alpha^D_ 2 & 0 & \alpha^D_3\\
\alpha^D_2 & \alpha^D_3 & 0
\end{array}
\right) \, \epsilon \langle h_u \rangle \, .
\end{equation}
As one can see, ${\cal M}^D_\nu$ receives at LO only contributions from flavons whose VEVs preserve $Z_2^u$.
Thus, the matrices ${\cal M}_R$ and ${\cal M}^D_\nu$ are invariant under different $Z_2$ subgroups of $D_{14}$
and, as a consequence, the light neutrino mass matrix $\mathcal{M}_\nu$ is not invariant under any of these two symmetries.
We can parametrize it as
\begin{equation}
\label{eq:mlight_LO}
\mathcal{M}_\nu = \left( \begin{array}{ccc}
 x & z+ s \sin\gamma & z - s \sin\gamma\\
 z+s \sin\gamma &  (u+y) + 2 \, p \sin\gamma & (u-y)\\
 z-s \sin\gamma & (u-y) & (u+y) -2 \, p \sin\gamma
\end{array}
\right) \, \frac{\epsilon \, \langle h_u \rangle^2}{\Lambda}
\end{equation}
with $\sin\gamma \approx \lambda$ and complex parameters $u$, $x$, $y$, $z$ and $s$, $p$. Applying a rotation with $\theta=\pi/4$ in the 2-3 sector we see that the third column and row
of $\mathcal{M}_\nu$ become proportional to
\begin{equation}
\left( \begin{array}{c}
\sqrt{2} \, s \sin\gamma \\ 2 \, p \sin\gamma \\ 2 \, y
\end{array} \right)
\end{equation}
showing that 
\begin{equation}
\theta_{23}^\nu - \pi/4 \sim p/y \, \sin\gamma \approx \lambda \;\;\ \mbox{and} \;\;\; \theta_{13}^\nu \sim s/( \sqrt{2} \, y ) \, \sin\gamma \approx \lambda \, .
\end{equation}
 Note that these deviations from the $\mu\tau$ symmetric result
are unrelated, because the one of $\theta_{23}^\nu$ is determined by the parameter $p$ and the one of $\theta_{13}^\nu$ by $s$. Defining
\begin{equation}
\zeta = |x|^2 - 4 |u|^2 \;\;\; , \;\;\; 
\kappa = 8 |z|^2 \left| \bar{x} \, \mathrm{e}^{2 i \alpha_z}  + 2 \, u \right|^2
\end{equation}
with $\alpha_{z}$ being the phase of the complex parameter $z$, we find for the mixing angle $\theta_{12}^\nu$ in zeroth order in the  
expansion parameter $\sin\gamma \approx \lambda$
\begin{equation}
\sin^2 \theta_{12}^\nu = \frac{1}{2} \left( 1 + \frac{\zeta}{\sqrt{\zeta^2 + \kappa}} \right) 
\end{equation}
which is naturally large (for non-hierarchical parameters).
The light neutrino mass $m_3$ is at lowest order driven by the parameter $y$, $m_3 \propto 2 \, |y|$, while the other two masses $m_{1,2}$ are determined by $u$, $x$ and $z$
\begin{equation}
\Delmsol \equiv m_2^2 - m_1^2 \propto \sqrt{\zeta^2 + \kappa} \;\;\; , \;\;\;  m_1^2 + m_2^2 \propto  |x|^2 + 4 \left( |u|^2 + |z|^2 \right) \, .
\end{equation}
In the case of normal hierarchy, we see that a small value of the ratio $r$ of the solar and the
atmospheric mass square differences, $r= \Delmsol/\Delmatm$ ($\Delta \mathrm{m}^2_{ij} \equiv m^2_i -m^2_j$), is easily achieved for $y$ being a factor 3 larger than the other parameters. 
On the other hand, an inversely ordered light neutrino mass spectrum can be, for example, achieved for small $y$ and larger $z$ (ensuring $m_3 < m_{1,2}$)
and $\zeta$ and $\kappa$ small (ensuring $\Delmsol$ and hence $r$ small).
Since the mixing angles $\theta_{ij}^e$ in the charged lepton sector are smaller or of order $\epsilon \approx \lambda^2$, see Eq.(\ref{chmix}), the angles $\theta_{ij}^\nu$
determine at LO the values of the lepton mixing angles.
Let us stress again that the preservation of different $Z_2$ subgroups of $D_{14}$ through the Majorana mass matrix of the right-handed neutrinos and the
Dirac neutrino mass matrix, respectively, is crucial for the achievement of $\theta_{13}^l \approx \mathcal{O}(\lambda)$. 
The relation between the size of the Cabibbo angle and the deviation from $\mu\tau$ symmetric mixing is not an accident, but based on the fact that the symmetry $Z_2^d$
determines the structure of the Majorana mass matrix of the right-handed neutrinos as well as of the down quark mass matrix and $Z_2^u$ the form of the Dirac neutrino
mass matrix and of the up quark mass matrix. For further details see \Secref{sec:D14sub}.

We can estimate the size of the cutoff scale $\Lambda$  using \Eqref{eq:mlight_LO}.
For an absolute neutrino mass scale $m_0$ of order $0.1 \eV$, $\langle h_u \rangle \approx 100$ GeV and $\epsilon \approx 0.04$ we get 
\begin{equation}
\label{eq:Lambda_estimate}
\La \simeq  \frac{\epsilon \langle h_u \rangle^2}{m_0} \simeq 4\cdot 10^{12}\, \GeV\,.
\end{equation}
We note that the contribution from the type I seesaw mechanism dominates over the one from the Weinberg operator, if the cutoff
scale associated with the latter is also $\La$ (or even larger). As we discuss in \Secref{sec:Weinberg}, the leading contribution of the Weinberg operator
stems from two flavon insertions so that we estimate its size to be of order $\epsilon^2 \langle h_u \rangle^2/\La$
and it is thus suppressed by a factor $\epsilon$ with respect to the leading term coming from the type I seesaw mechanism, see \Eqref{eq:mlight_LO}.

\subsubsection{NLO results}

At the NLO level several additional operators induce corrections to the mass matrix $\mathcal{M}_{R}$. At the level
of two flavon insertions, one up-type and one down-type one, the terms
\begin{equation}
\frac{1}{\La} \nu_1^c (\nu_D^c\psi^u\chi^d) + \frac{1}{\La} \nu_1^c (\nu_D^c\psi^u) \sigma + \frac{1}{\La} \nu_1^c (\nu_D^c\chi^u\psi^d) 
\end{equation}
contribute to the (12) and (13) elements of $\mathcal{M}_{R}$. At the same level also the shifts of the VEVs contribute, if they are of the generic size
$\epsilon \times \mbox{VEV}$
\begin{equation}
\nu^c_1 (\nu^c_D \delta\psi^d) \, .
\end{equation}
The elements (22) and (33) receive contributions from the operators
\begin{equation}
\frac{1}{\La} (\nu_D^c\psi^u)(\nu_D^c\psi^d) + \frac{1}{\La}(\nu_D^c\nu_D^c\chi^u\xi^d) 
+ \frac{1}{\La}(\nu_D^c\nu_D^c\chi^u) \sigma+ \frac{1}{\La}(\nu_D^c\nu_D^c\xi^u\chi^d) 
\end{equation}
and a contribution from plugging in the shifted VEV of  $\chi^d_{1,2}$
\begin{equation}
(\nu^c_D \nu^c_D \delta\chi^d) \, .
\end{equation}
Corrections to the non-zero elements (11), (23) and (32) of $\mathcal{M}_R$  from operators containing two flavons are
 absorbed into the LO parameters $\alpha^M_{1,2}$. 
Thus, the most general form of $\mathcal{M}_R$ can be parametrized as
\begin{equation}
\mathcal{M}_R = \left( \begin{array}{ccc}
 \alpha^M_1 & \alpha^M_3 \, \mathrm{e}^{i \gamma}  & \alpha^M_3 \, \mathrm{e}^{-i \gamma} + \beta^M_1 \epsilon\\
  \alpha^M_3  \, \mathrm{e}^{i \gamma}  &  \alpha^M_4 \,  \, \mathrm{e}^{2 i \gamma} & \alpha^M_2\\
  \alpha^M_3  \, \mathrm{e}^{-i \gamma} + \beta^M_1 \epsilon & \alpha^M_2 &  \alpha^M_4 \,  \, \mathrm{e}^{-2 i \gamma} + \beta^M_2 \epsilon\\
\end{array}
\right) \, \epsilon \, \Lambda
\end{equation}
with $\alpha^M_{i}$ and $\beta^M_{i}$ being complex numbers with absolute values of order one. The parameters $\alpha^M_i$ are re-defined in such a way to
absorb some of the subleading corrections.

Also the Dirac neutrino mass matrix $\mathcal{M}^D_\nu$ acquires a more general form, if subleading terms are included.
Clearly, all corrections to the (11) element can be absorbed into the leading contribution, so we do not list operators contributing in this way.
Plugging in the shifted VEVs into the operators in Eq.(\ref{Diracnu_LO}),
\begin{equation}
 \frac{1}{\La} (L_D\delta \chi^u) \nu^c_1 h_u + \frac{1}{\La} (L_D \nu^c_D \delta\psi^u) h_u \, ,
\end{equation}
leads to corrections of relative order $\epsilon$, if all VEV shifts are of order $\epsilon$ in units of the generic flavon VEV. They disturb
 the tight relations between the (21) and (31) elements as well as the (23) and (32) elements at relative order $\epsilon$.
At the level of two flavon insertions we only find operators containing up-type flavons, 
\begin{equation}
\frac{1}{\La^2} (L_1 \eta^u)(\nu_D^c\psi^u) h_u  + \frac{1}{\La^2} (L_1\nu_D^c\chi^u\xi^u) h_u
+ \frac{1}{\La^2} (L_1\nu_D^c\xi^u\xi^u) h_u
\end{equation}
which give rise to non-zero and equal (12) and (13) elements of $\mathcal{M}^D_\nu$ , if the LO VEVs are plugged in. 
Taking into account VEV shifts as well the latter operators induce a relative difference of order $\epsilon$
in the (12) and (13) elements. Similarly, equal (22) and (33) elements are generated with operators of the form
(some of them stand for more than one inequivalent $D_{14}$ contraction leading to distinct contributions to the Dirac neutrino mass matrix) 
\begin{equation}
\frac{1}{\La^2} (L_D \nu^c_D \psi^u\chi^u) h_u + \frac{1}{\La^2} (L_D \nu^c_D \psi^u\xi^u) h_u + \frac{1}{\La^2} (L_D \nu^c_D \xi^u\eta^u) h_u \, ,
\end{equation}
if LO VEVs are plugged in. If we take into account the shifted VEVs we find deviations of relative order $\epsilon$ from the equality of 
the (22) and (33) elements. 
The (21) and (31) elements as well as the (23) and (32) elements also receive contributions from operators
with two flavons whose effect however can be absorbed into the leading terms. 
Eventually, operators  with three flavon insertions, two down-type flavons and
the fields $\chi^e_{1,2}$, have to be considered, since also they contribute to the perturbation of the equality of the (12) and (13) elements 
and the (22) and (33) elements. Relevant for the former elements are the operators
\begin{eqnarray}\nonumber
&&\frac{1}{\La^3} (L_1 \nu^c_D  \psi^d \eta^d \chi^e)  h_u + \frac{1}{\La^3} (L_1 \nu^c_D \chi^d \chi^d \chi^e) h_u 
+ \frac{1}{\La^3} (L_1 \nu^c_D  \chi^d \xi^d \chi^e)  h_u 
\\
&&+ \frac{1}{\La^3} (L_1 \nu^c_D \xi^d \xi^d \chi^e) h_u
+ \frac{1}{\La^3} (L_1 \nu^c_D  \xi^d \chi^e) \sigma  h_u  
\, .
\end{eqnarray}
Operators of this type also induce deviations from the equality of the (22) and (33) elements of relative order $\epsilon$
(again, some of them represent more than one inequivalent $D_{14}$ contraction):
\begin{eqnarray}\nonumber
&&\frac{1}{\La^3} (L_D \nu^c_D  \psi^d \chi^d \chi^e) h_u + \frac{1}{\La^3} (L_D \nu^c_D \psi^d \xi^d \chi^e) h_u
+ \frac{1}{\Lambda^3} (L_D \nu^c_D  \psi^d  \chi^e) \sigma h_u 
\\
&&+ \frac{1}{\Lambda^3} (L_D \nu^c_D  \xi^d \eta^d \chi^e) h_u+ \frac{1}{\Lambda^3} (L_D \nu^c_D  \chi^d \eta^d \chi^e) h_u\,.
\end{eqnarray}
The other elements of $\mathcal{M}^D_\nu$ are affected by this type of operators as well, however, their effect can be absorbed into contributions from 
operators with less flavons.  The same is true for the second class of operators with three flavons, this time up-type fields. 
So, the most general form of the Dirac neutrino mass matrix can be parametrized as
\begin{equation}
\mathcal{M}^D_\nu = \left( \begin{array}{ccc}
 \alpha^D_1 & \alpha^D_4 \epsilon & \alpha^D_4 \epsilon + \beta^D_3 \epsilon^2\\
\alpha^D_ 2 & \alpha^D_5 \epsilon & \alpha^D_3 + \beta^D_2 \epsilon\\
\alpha^D_2 + \beta^D_1 \epsilon & \alpha^D_3 & \alpha^D_5 \epsilon + \beta^D_4 \epsilon^2 
\end{array}
\right) \, \epsilon \langle h_u \rangle
\end{equation}
with $\alpha^D_i$ and $\beta^D_i$ being complex parameters with absolute value of order one.
The parameters $\alpha^D_i$ are equal to the parameters in the LO result only up to corrections of order $\epsilon$.
The parametrization of the light neutrino mass matrix shown in Eq.(\ref{eq:mlight_LO}) is already the most general one, we can achieve in our setup, and all NLO contributions can be
captured by re-defining the complex parameters $u$, $x$, $y$, $z$ and $s$, $p$.

\subsubsection{Contributions from the Weinberg operator}
\label{sec:Weinberg}

As mentioned, the Weinberg operator contains at least two flavons. Four different operators can be found
at this level (one being a down-type flavon and the other one being $\chi^e_{1,2}$)
\begin{equation}
\label{eq:Weinberg}
\frac{1}{\La^3} (L_1 L_1) (\chi^e \chi^d) h_u^2 +  \frac{1}{\La^3} (L_1 L_D \chi^e \eta^d) h_u^2
+ \frac{1}{\La^3} (L_D\,L_D) (\chi^e \chi^d) h_u^2 + \frac{1}{\La^3} (L_D \chi^e) (L_D \chi^d) h_u^2
\end{equation}
generating the (11), $(i3)$, $(3i)$ entries of the mass matrix $\mathcal{M}^W$, associated with the Weinberg operator.
Non-zero (12) and (21) entries arise, if the shifted VEVs are plugged into the second operator in \Eqref{eq:Weinberg}, and the (22) element becomes non-zero, if the 
shifted VEVs are plugged into the forth operator in \Eqref{eq:Weinberg}.
Apart from that operators with three flavons, one up-type, one down-type flavon and the fields $\chi^e_{1,2}$, also contribute
at this level to the (12), (21) and (22) elements (and lead to subleading contributions to the other matrix elements). 
 Operators, potentially relevant for the generation of the (12) and (21) elements, are
\begin{eqnarray}
&& \frac{1}{\La^4} (L_1 L_D  \psi^u\chi^d \chi^e) h_u^2 
+  \frac{1}{\La^4} (L_1 L_D  \psi^u \xi^d \chi^e) h_u^2
+  \frac{1}{\La^4} (L_1 L_D  \chi^u \psi^d \chi^e) h_u^2 \nonumber
\\  
&+& \frac{1}{\La^4} (L_1 L_D  \xi^u \psi^d \chi^e) h_u^2
+ \frac{1}{\La^4} (L_1 L_D  \xi^u \eta^d \chi^e) h_u^2 
+  \frac{1}{\La^4} (L_1 \eta^u)(L_D \xi^d \chi^e) h_u^2  \nonumber
\\  
&+&  \frac{1}{\La^4} (L_1 \eta^u)(L_D \chi^e) \sigma h_u^2\,.
\end{eqnarray}
And similarly, those which can contribute to the (22) element, read
\begin{eqnarray} \nonumber
&& \frac{1}{\La^4} (L_D \chi^e)(L_D  \psi^u \psi^d) h_u^2 
+ \frac{1}{\La^4} (L_D L_D \psi^u \eta^d \chi^e) h_u^2
+ \frac{1}{\La^4} (L_D L_D \xi^d)(\chi^u \chi^e) h_u^2 
\\ \nonumber
&+& \frac{1}{\La^4} (L_D \chi^u)(L_D \chi^e \xi^d) h_u^2
+ \frac{1}{\La^4} (L_D \chi^u)(L_D \chi^e) \sigma h_u^2
+  \frac{1}{\La^4} (L_D L_D  \xi^u)(\chi^d \chi^e) h_u^2 
\\ 
&+& \frac{1}{\La^4} (L_D \chi^d)(L_D \chi^e \xi^u) h_u^2
+  \frac{1}{\La^4} (L_D L_D  \xi^u \xi^d \chi^e) h_u^2
+  \frac{1}{\La^4} (L_D L_D  \eta^u \psi^d \chi^e) h_u^2 \; .
\end{eqnarray}
Hence, the most general form of the matrix $\mathcal{M}^W$ is
\begin{equation}
\mathcal{M}^W =  \left( \begin{array}{ccc}
\alpha_1^W & \beta_1^W \epsilon& \alpha_2^W\\
\beta_1^W \epsilon & \beta_2^W \epsilon  & \alpha_3^W\\
\alpha_2^W &  \alpha_3^W  & \alpha_4^W
\end{array}
\right) \, \frac{\epsilon^2 \langle h_u \rangle^2}{\La}
\end{equation}
with all parameters $\alpha^W_i$ and $\beta^W_i$ being complex numbers with absolute value of order one. As one can check, the contribution $\mathcal{M}^W$
to the light neutrino mass matrix can be absorbed into the form given in Eq.(\ref{eq:mlight_LO}) by a simple re-definition of the complex parameters $u$, $x$, $y$, $z$ and $s$, $p$.

\section{Quark sector}
\label{sec:quarks}

We briefly discuss the results for the quark sector putting emphasis on the differences to the
original setup \cite{D14_quarks}.
The operators, contributing to the down quark mass matrix, at the leading as well as the subleading level coincide with those obtained in \cite{D14_quarks} and
are not repeated in detail here. In general, we find at LO operators with one flavon $\{\psi^{d}_{1,2}, \chi_{1,2}^{d}, \xi^{d}_{1,2}, \eta^{d}, \sigma\}$ 
whose leading VEVs conserve the group $Z_2^d$, whereas subleading operators, involving one up-type and one down-type flavon, break this remnant $Z_2$ symmetry. 
Including leading and subleading terms as well as contributions attributed to the shifts in the flavon VEVs, 
the down quark mass matrix $\mathcal{M}_d$ can be written as
\begin{equation}
\label{eq_Md_NLO}
\mathcal{M}_d = \left(
\begin{array}{ccc}
	 \beta^d_1 \, t \, \epsilon^2 & t \, (\alpha^d_1 \, \epsilon + \beta^d_4 \, \epsilon^2) & \beta^d_5 \, \epsilon^2\\
	 \beta^d_2 \, t \, \epsilon^2 & \alpha^d_1 \, \mathrm{e} ^{- 2 i \gamma} \, t \, \epsilon & \beta^d_6 \, \epsilon^2\\
	 \beta^d_3 \, t \, \epsilon^2 & \alpha^d_2 \, t \, \epsilon & y_b \, \epsilon 
\end{array}
\right) \, \langle h_d \rangle 
\end{equation}
with complex parameters  $\alpha^d_i$, $\beta^d_i$ and $y_b$. This result is in accordance with the findings of \cite{D14_quarks}.

In the case of the up quarks we see that all operators with up to two flavons coincide with those found in the model in \cite{D14_quarks}. 
If the VEV shifts are neglected, the group $Z_2^u$ is preserved in the up quark sector up to this level.
Corrections arise, as in \cite{D14_quarks}, through VEV shifts as well as operators with three
flavons. The latter can be divided into two classes: $Z_2^u$ symmetry preserving operators, with three up-type flavons, and $Z_2^u$ symmetry breaking ones.
 The former contributions are again the same as in \cite{D14_quarks}. However, the operators with three flavons, which
induce $Z_2^u$ breaking contributions, have another form, due to the extension of the symmetry, which segregates the different sectors,
 from $Z_3$ to $Z_7$. They contain 
two down-type flavons and the fields $\chi^e_{1,2}$. Relevant contributions are due to the operators
\begin{eqnarray}
&& \frac{\theta^2}{\La^5}  Q_3 (u^c \psi^d \xi^d \chi^e) h_u + \frac{\theta^2}{\La^5} Q_3 (u^c \eta^d) (\chi^d \chi^e) h_u
\label{eq:line1}
\\ 
&& \frac{1}{\La^3} (Q_D c^c \psi^d \eta^d \chi^e) h_u + \frac{1}{\La^3} (Q_D c^c \chi^d \chi^d \chi^e) h_u
+ \frac{1}{\La^3} (Q_D c^c \chi^d \xi^d \chi^e) h_u 
\nonumber
\\ \label{eq:line2}
&&+ \frac{1}{\La^3} (Q_D c^c \xi^d \xi^d \chi^e) h_u
+ \frac{1}{\La^3} (Q_D c^c \xi^d \chi^e) \sigma h_u
\\ 
&& \frac{\theta^2}{\La^5} (Q_D \psi^d \chi^e)(u^c \eta^d) h_u + \frac{\theta^2}{\La^5} (Q_D u^c \chi^d \chi^d \chi^e) h_u
+ \frac{\theta^2}{\La^5} (Q_D u^c \chi^d \xi^d \chi^e) h_u 
\nonumber
\\ \label{eq:line3} 
&&+ \frac{\theta^2}{\La^5} (Q_D u^c \xi^d \xi^d \chi^e) h_u
+ \frac{\theta^2}{\La^5} (Q_D u^c \xi^d \chi^e) \sigma h_u \, .
\end{eqnarray}
As one can see, the operators in \Eqref{eq:line1} generate the (31) element of the up quark mass matrix $\mathcal{M}_u$, while the operators in \Eqref{eq:line2} 
give rise to deviations from the equality of the (12) and (22) elements. Similarly, operators given in \Eqref{eq:line3} break the close relation of the (11) and
(21) elements coming from the presence of the group $Z_2^u$ in the up quark sector. 
$\mathcal{M}_u$ can thus be cast into the form
\begin{equation}
\label{eq:Mu_NLO}
\mathcal{M}_u = \left(
\begin{array}{ccc}
	t^2 \, (-\alpha^u_1 \, \epsilon^2 + \beta_1^u \, \epsilon^3) & \alpha^u_2 \, \epsilon^2 + \beta_2^u \, \epsilon^3 
				& \alpha^u_3 \, \epsilon + \beta_3^u \, \epsilon^2\\
	\alpha^u_1 \, t^2 \, \epsilon^2 & \alpha^u_2 \, \epsilon^2 & \alpha^u_3 \, \epsilon\\
	\beta^u_4 \, t^2 \, \epsilon^3 & \alpha^u_4 \, \epsilon & y_t 
\end{array}
\right) \, \langle h_u \rangle 
\end{equation}
with all parameters being complex. The structure of ${\cal M}_u$ is the same as in the original setup \cite{D14_quarks}.
As a consequence, the results of quark masses and mixing parameters are the same. We briefly summarize these, taking
$t \approx \epsilon$
\begin{eqnarray}\nonumber
&&m_u: m_c: m_t \sim \epsilon^4: \epsilon^2: 1 \;\;\; \mbox{with} \;\;\; m_t \approx |y_t|  \langle h_u \rangle \, , \\
\label{eq:masses_NLO}
&&m_d: m_s: m_b \sim \epsilon^2: \epsilon: 1 \;\;\;\;\, \mbox{with} \;\;\; m_b \approx |y_b|  \langle h_d \rangle \epsilon
\end{eqnarray}
and for the elements of the quark mixing matrix and the Jarlskog invariant $J_{CP}$ \cite{jarlskog}
we get
\begin{eqnarray}\nonumber
\vert V_{ud} \vert, \vert V_{cs} \vert  = \cos \gamma + \mathcal{O}(\epsilon) \approx  0.97 \, , \,\,
&& \vert V_{cb} \vert, \vert V_{ts} \vert, \vert V_{ub} \vert, \vert V_{td} \vert   \sim \mathcal{O}(\epsilon) \, , \,\,\\
\vert V_{us} \vert, \vert V_{cd} \vert = \sin \gamma + \mathcal{O}(\epsilon) \approx  0.22 \, , \,\,
&&  \vert V_{tb} \vert \approx 1 +  \mathcal{O}(\epsilon^2)  \, , \,\, J_{CP} \sim  \mathcal{O}(\epsilon^2) \, .
\end{eqnarray}
As commented in \cite{D14_quarks}, using the freedom of the parameters in $\mathcal{M}_d$ and $\mathcal{M}_u$,
the best fit values of all masses and mixing parameters can be accommodated without tuning. The seemingly 
too large value of the Jarlskog invariant $J_{CP}$ is suppressed by a numerical factor 0.11.

\section{Flavon superpotential}
\label{sec:flavons}

In order to construct the flavon superpotential $w_f$ we add two ingredients (see e.g. \cite{AF2}): 
a set of so-called driving fields whose $F$-terms account for the vacuum alignment of the flavon fields
and an $R$-symmetry $U(1)_R$ under which
matter fields have charge $+1$, flavon fields, $h_{u,d}$ and $\theta$ are neutral and driving fields have charge $+2$. 
In this way, all terms in the superpotential $w_f$ are linear in the driving fields and at the same time
these fields are not involved in operators contributing directly to fermion masses. 
Since we expect the flavor symmetry to be broken at high energies, soft supersymmetry 
breaking effects are not relevant in the discussion of the vacuum alignment of the flavons.
We divide the flavon superpotential into three parts
\begin{equation}
w_f = w_{f,l} +  w_{f,u} + w_{f,d} \; .
\end{equation}
In the following we first discuss the LO form of $w_{f,l}$, $w_{f,u}$ and $w_{f,d}$, and then we present the terms contributing at NLO
to $w_f$ and their effect on the LO vacuum alignment.

\subsection{Lepton sector}
\label{subsec:flavons_l}

The superpotential $w_{f,l}$ responsible for the vacuum alignment of $\chi^e_{1,2}$ has
a very simple form, because we only need to introduce the
driving field $\sigma^{0e}$ which transforms trivially under $D_{14}$ and acquires a phase $\omega_7^3$
under $Z_7$. It is, as the other driving fields, see \Tabref{tab:driving}, neutral under the FN symmetry. 
$w_{f,l}$ reads
\begin{equation}
w_{f,l} = a_l \, \sigma^{0e} \, \chi^e_1 \, \chi^e_2 \; .
\end{equation}
The $F$-term of $\sigma^{0e}$ only allows
vacua in which at least one of the two fields $\chi^e_{1,2}$ has a vanishing VEV, i.e. the trivial vacuum and the two solutions shown in \Eqref{eq:align_che}.
In the latter case, the VEV $v^e$ is a free parameter. It is interesting to note 
that both vacua, shown in \Eqref{eq:align_che}, break $D_{14}$ in the charged lepton sector to the $Z_2$ subgroup
which is generated by the element $\mathrm{A}^7$, since $\chi^e_{1,2}$ transform as unfaithful representation of $D_{14}$, see \Secref{sec:D14sub}.
This way of aligning the vacuum is similar to the one found in \cite{S3_alt}.
\begin{table}
\begin{center}
\begin{tabular}{|c||c||c|c|c|c||c|c|c|c|}\hline
Field & $\sigma^{0e}$ & $\sigma^{0u}$ & $\psi^{0u}_{1,2}$ & $\varphi^{0u}_{1,2}$ & $\rho^{0u}_{1,2}$ & 
 $\sigma^{0d}$ & $\psi^{0d}_{1,2}$ & $\varphi^{0d}_{1,2}$ & $\rho^{0d}_{1,2}$\\ 
\hline
\Groupname{D}{14}& $\MoreRep{1}{1}$ & $\MoreRep{1}{1}$ & $\MoreRep{2}{1}$ & $\MoreRep{2}{3}$ & $\MoreRep{2}{5}$ & 
$\MoreRep{1}{1}$ & $\MoreRep{2}{1}$ & $\MoreRep{2}{3}$  & $\MoreRep{2}{5}$\\
\Groupname{Z}{7} &  $\omega_7^3$ & $1$ & $1$ & $1$ & $1$  & $\omega_7^2$ & $\omega_7^2$ & $\omega_7^2$ & $\omega_7^2$\\ 
\hline
\end{tabular}
\end{center}
\begin{center}
\normalsize
\begin{minipage}[t]{12cm}
\caption[Driving Fields of the Model]{{\it Driving fields of the model and 
their transformation properties under the flavor 
symmetry $D_{14} \times Z_{7}$. Similar to the flavons none of the
driving fields is charged under $U(1)_{FN}$. The fields indexed with a $u$ ($d$, $e$)
drive the VEVs of the flavons giving  masses to the up (down) quarks (charged leptons) at lowest order.
Note that all these fields have a $U(1)_R$ charge $+2$.}
\label{tab:driving}}
\end{minipage}
\end{center}
\end{table}
%

\subsection{Quark sector}
\label{subsec:flavons_q}

In the construction of the superpotentials $w_{f,u}$ and $w_{f,d}$ we closely follow \cite{D14_quarks} and thus introduce the same driving
fields, listed for convenience in \Tabref{tab:driving}, only changing 
the charge of the down-type driving fields under the auxiliary cyclic symmetry appropriately. As a consequence, we find the same terms in the superpotential
at the renormalizable level and thus also the same results for the vacuum alignment. 
We note that in \cite{D14_quarks} the mass term, $\sigma^{0u} \left( M^u_\sigma \right)^2$, has been forgotten, which however has no relevant impact.
For the sake of completeness, we display the correct superpotential $w_{f,u}$ 
\begin{eqnarray}\label{eq:spup}\nonumber
w_{f,u} =& & M_{\psi}^u \left(\psi_1^u \psi_2^{0u}+\psi_2^u \psi_1^{0u}\right) + a_u  \left(\psi_1^u \chi_1^u \varphi_2^{0u}+\psi_2^u \chi_2^u \varphi_1^{0u}\right) 
+ b_u \left(\psi_1^u \chi_2^u \psi_1^{0u}+\psi_2^u \chi_1^u \psi_2^{0u}\right) \\
&+& c_u \left(\psi_1^u \xi_2^u \varphi_1^{0u}+\psi_2^u \xi_1^u \varphi_2^{0u}\right) 
+ d_u \eta^u \left(\xi_1^u \varphi_1^{0u} + \xi_2^u \varphi_2^{0u}\right) + e_u \left(\psi_1^u \xi_1^u \rho_2^{0u} + \psi_2^u \xi_2^u \rho_1^{0u}\right) \nonumber\\
&+& f_u \eta^u \left(\chi_1^u \rho_1^{0u}+\chi_2^u \rho_2^{0u}\right)+ g_u\,\sigma^{0u}\psi_1^u \psi_2^{u}+l_u\sigma^{0u}\chi_1^u \chi_2^{u}+
n_u\sigma^{0u}\xi_1^u \xi_2^u  \nonumber \\ &+& q_u \sigma^{0u}(\eta^u)^2  + \sigma^{0u} \left( M^u_\sigma \right)^2\,.
\end{eqnarray}
The conditions for the vacuum alignment are given by the $F$-terms and are the same as shown in \cite{D14_quarks} apart from the one 
associated with the driving field $\sigma^{0u}$ which reads
\begin{equation}
\frac{\partial w_f}{\partial \sigma^{0u}} = g_u \psi_1^u \psi_2^{u}+l_u\chi_1^u \chi_2^{u}+
n_u\xi_1^u \xi_2^u   + q_u (\eta^u)^2  +  \left(M^u_\sigma\right)^2 = 0 \;.  
\end{equation}
Solving the equations associated with the $F$-terms of the up-type driving fields, we get as unique solution for the VEVs of the up-type flavons
(excluding solutions which require some VEVs to vanish or some of the parameters in the flavon superpotential to be zero)
\begin{equation}
\label{eq:vac_u_LO}
\!\!\!\!\!\!\! \!\!\!\!\left( \begin{array}{c} \VEV{\psi_1^u} \\ \VEV{\psi_2^u} \end{array} \right) = v^u \left( \begin{array}{c} 
\mathrm{e}^{-2 i \gamma k_u} \\ 1 \end{array} \right) , \;
\left( \begin{array}{c} \VEV{\chi_1^u} \\ \VEV{\chi_2^u} \end{array} \right) = w^u \, \mathrm{e}^{2 i \gamma k_u}  
\left( \begin{array}{c} \mathrm{e}^{-4 i \gamma k_u}  \\  1 \end{array} \right) , \;
\left( \begin{array}{c} \VEV{\xi_1^u} \\ \VEV{\xi_2^u} \end{array} \right) = z^u \, \mathrm{e}^{4 i \gamma k_u} 
\left( \begin{array}{c} \mathrm{e}^{-8 i \gamma k_u} \\  1 \end{array} \right)
\end{equation}
with
\begin{eqnarray}
\label{eq:value_wu_zu}
w^u  &=&  -\frac{M_{\psi}^u}{b_u} \; , \;\; 
z^u  =  \frac{w^u}{2  d_u e_u} \left( c_u f_u \pm \sqrt{4 a_u d_u e_u f_u+ (c_u f_u)^2} \right) \nonumber \\
v_u^2 &=&  - \frac{(M^u_\sigma)^2 + l_u (w^u)^2 + n_u (z^u)^2}{g_u + q_u \left(\frac{e_u z^u}{f_u w^u}\right)^2} \, \mathrm{e}^{2 i \gamma k_u}
\; , \;\; 
 \VEV{\eta^{u}} =- \frac{e_u}{f_u} \frac{v^u z^u}{w^u} \, \mathrm{e}^{-8 i \gamma k_u} \,,
\end{eqnarray}
and $k_u= 0, 2, ..., 12$. We can set $k_u=0$, compare \Eqref{eq:vacuumup_LO}, because, as has been argued in \cite{D14_quarks}, the Cabibbo angle
depends on the difference of $k_u$ and $k$ ($k$ is the parameter appearing in the LO vacuum of the down-type flavons, see \Eqref{eq:vacuumdown_LO}), 
so that the value of $k_u$ itself is not relevant. Note that the VEVs of all up-type flavons are
fixed in terms of the two mass parameters $M^u_{\psi}$ and $M^u_{\sigma}$.

$w_{f,d}$ has the same structure as in \cite{D14_quarks} and leads
to the vacuum shown in \Eqref{eq:vacuumdown_LO}. The VEVs of the down-type flavons fulfill relations very similar to the ones given in \Eqref{eq:value_wu_zu}, see also \cite{D14_quarks}, with
$\langle \sigma \rangle$ being a free parameter.

\subsection{NLO corrections}
\label{sec:flavons_nlo}

The NLO corrections to the flavon superpotential $w_f$ arise at the non-renormalizable level from terms with one driving field
and three flavons. At this level the separation between the different symmetry breaking sectors is lost and terms containing for example
an up-type driving field and the fields $\chi^e_{1,2}$ show up. As a consequence the vacuum, presented in Eqs. (\ref{eq:vacuumup_LO},\ref{eq:vacuumdown_LO},\ref{eq:align_che}), gets shifted.
We discuss which terms arise at this level and show that the size of the shifts is $\epsilon$ in units of the generic flavon VEV.

\subsubsection{Lepton sector}
\label{sec:flavons_l_nlo}

Since the structure of $w_{f,l}$ aligning the VEV of $\chi^e_{1,2}$ is very simple, also the
NLO corrections which induce a shift to the vacuum of $\chi^e_{1,2}$ have a simple form. The latter is parametrized as 
\begin{equation}
\left( \begin{array}{c} \VEV{\chi_1^e} \\ \VEV{\chi_2^e}\end{array} \right)
= \left( \begin{array}{c} v^e \\ \delta v^e \end{array} \right) 
\end{equation}
and the free parameter $v^e$ is not determined by the NLO corrections. 
There are two types of NLO corrections: either two of the three flavons are $\chi^e_{1,2}$ and the third
one is an up-type flavon or all three flavons belong to the set $\{ \psi^{d}_{1,2}, \chi_{1,2}^{d}, \xi^{d}_{1,2}, \eta^{d}, \sigma\}$.
Only one contribution belongs to the first category, while the second one contains seven
terms. The corrections to $w_{f,l}$ can be summarized as
\begin{equation}
\Delta w_{f,l} = \frac{1}{\La} \, \sum _{k=1} ^{8} p^l_k \, I^{P,l}_{k}
\end{equation}
with the invariants $I^{P,l}_{k}$ defined as 
\begin{equation}
\begin{array}{ll}
I_1^{P,l} = \sigma^{0e} \left( \left(\chi^e_1 \right)^2 \xi^u_2 +  \left(\chi^e_2 \right)^2 \xi^u_1 \right) &
I_5^{P,l} = \sigma^{0e} \sigma \left( \eta^d \right)^2 \\
I_2^{P,l} = \sigma^{0e} \sigma \psi^d_1 \psi^d_2 &
I_6^{P,l} = \sigma^{0e} \sigma^3\\
I_3^{P,l} = \sigma^{0e} \sigma \chi^d_1 \chi^d_2 &
I_7^{P,l} = \sigma^{0e} \left( \left(\psi^d_1 \right)^2 \chi^d_2 + \left(\psi^d_2 \right)^2 \chi^d_1 \right) \\
I_4^{P,l} = \sigma^{0e} \sigma \xi^d_1 \xi^d_2 &
I_8^{P,l} = \sigma^{0e} \left( \left(\chi^d_1 \right)^2 \xi^d_2 + \left(\chi^d_2 \right)^2 \xi^d_1 \right) \; .
\end{array}
\end{equation}
Computing the size of the shift $\delta v^e$, we find it to be generically of the order $\epsilon^2 \, \Lambda$, i.e. relatively suppressed to the leading 
VEV $v^e$ by a factor $\epsilon$.

\subsubsection{Quark sector}
\label{sec:flavons_q_nlo}

Concerning the superpotential $w_{f,d}$ driving the VEVs  of the flavons $\{ \psi^{d}_{1,2}, \chi_{1,2}^{d}, \xi^{d}_{1,2}, \eta^{d}, \sigma\}$ we observe that all NLO 
terms have the same form as in the model \cite{D14_quarks}. Thus, the flavon VEVs receive the same shifts as in the model for quarks only.
In the case of $w_{f,u}$, the $Z_2^u$ symmetry preserving subleading terms which contain only up-type flavons
are the same as in \cite{D14_quarks}.
 The $Z_2^u$ symmetry breaking terms at the first non-renormalizable level instead differ, because
three down-type flavons cannot couple in a $Z_7$-invariant way to an up-type driving field. We find that combinations of two down-type
flavons and $\chi^e_{1,2}$ can couple to an up-type driving field
\begin{equation}
\Delta w_{f,u}^{\mathrm{add}} = \frac{1}{\La} \, \left( 
\sum \limits_{k=1}^3 p^J_k \, J^{P,u}_{k} + \sum \limits_{k=1}^5 r^J_k \, J^{R,u}_{k}
+  \sum \limits_{k=1}^4 s^J_k \, J^{S,u}_{k} +  \sum \limits_{k=1}^4 t^J_k \, J^{T,u}_{k} 
\right)
\end{equation}
with

\parbox{3in}{
\begin{eqnarray}\nonumber
J^{P,u}_1 &=& \sigma^{0u} \left( (\psi^d_1)^2 \chi^e_2 + (\psi^d_2)^2 \chi^e_1 \right) 
\\ \nonumber
J^{P,u}_2 &=& \sigma^{0u} \left( \chi^d_2 \xi^d_1 \chi^e_2 + \chi^d_1 \xi^d_2 \chi^e_1 \right)
\\ \nonumber
J^{P,u}_3 &=& \sigma^{0u} \sigma \left( \chi^d_1 \chi^e_2 + \chi^d_2 \chi^e_1 \right)
\end{eqnarray}}
\parbox{3in}{
\begin{eqnarray}\nonumber
J^{R,u}_1 &=& \left( \psi^{0u}_1 \psi^d_2 \chi^d_1 \chi^e_2 + \psi^{0u}_2 \psi^d_1 \chi^d_2 \chi^e_1 \right)
\\ \nonumber
J^{R,u}_2 &=& \left( \psi^{0u}_1 \psi^d_2 \chi^d_2 \chi^e_1 + \psi^{0u}_2 \psi^d_1 \chi^d_1 \chi^e_2 \right)
\\ \nonumber
J^{R,u}_3 &=& \left( \psi^{0u}_1 \psi^d_1 \xi^d_2 \chi^e_1 + \psi^{0u}_2 \psi^d_2 \xi^d_1 \chi^e_2 \right) 
\\ \nonumber
J^{R,u}_4 &=& \sigma \left( \psi^{0u}_1 \psi^d_1 \chi^e_2 + \psi^{0u}_2 \psi^d_2 \chi^e_1 \right)
\\ 
J^{R,u}_5 &=& \eta^d \left( \psi^{0u}_1 \xi^d_1 \chi^e_1 - \psi^{0u}_2 \xi^d_2 \chi^e_2 \right)
\end{eqnarray}}

and

\parbox{3in}{
\begin{eqnarray}\nonumber
J^{S,u}_1 &=& \left( \varphi^{0u}_1 \psi^d_1 \chi^d_2 \chi^e_2 + \varphi^{0u}_2 \psi^d_2 \chi^d_1 \chi^e_1 \right)
\\ \nonumber
J^{S,u}_2 &=& \left( \varphi^{0u}_1 \psi^d_2 \xi^d_2 \chi^e_1 + \varphi^{0u}_2 \psi^d_1 \xi^d_1 \chi^e_2 \right)
\\ \nonumber
J^{S,u}_3 &=& \sigma \left( \varphi^{0u}_1 \psi^d_2 \chi^e_2 + \varphi^{0u}_2 \psi^d_1 \chi^e_1 \right)
\\ \nonumber
J^{S,u}_4 &=& \eta^d \left( \varphi^{0u}_1 \chi^d_1 \chi^e_1 - \varphi^{0u}_2 \chi^d_2 \chi^e_2 \right)
\end{eqnarray}}
\parbox{3in}{
\begin{eqnarray}\nonumber
J^{T,u}_1 &=& \left( \rho^{0u}_1 \psi^d_2 \chi^d_2 \chi^e_2 + \rho^{0u}_2 \psi^d_1 \chi^d_1 \chi^e_1 \right)
\\ \nonumber
J^{T,u}_2 &=& \left( \rho^{0u}_1 \psi^d_1 \xi^d_2 \chi^e_2 + \rho^{0u}_2 \psi^d_2 \xi^d_1 \chi^e_1 \right)
\\ \nonumber
J^{T,u}_3 &=& \eta^d \left( \rho^{0u}_1 \xi^d_1 \chi^e_2 - \rho^{0u}_2 \xi^d_2 \chi^e_1 \right)
\\   
J^{T,u}_4 &=& \sigma \eta^d \left( \rho^{0u}_1 \chi^e_1 - \rho^{0u}_2 \chi^e_2 \right) \, .
\end{eqnarray}}

The shifted VEVs can be parametrized in the same way as in the original model \cite{D14_quarks}
\begin{eqnarray}
\nonumber
&& \!\!\!\!\! \!\!\!\!\! \VEV{\psi_i^u}  =  v^u +\delta v_i^u \; , \;\; 
\VEV{\chi_i^u}  =  w^u +\delta w_i^u \; , \;\;
\VEV{\xi_i^u}   =  z^u +\delta z_i^u \; , \;\;
\VEV{\eta^u}  =  -\frac{e_u}{f_u} \frac{v^u z^u}{w^u} + \delta \eta^u\\ \nonumber
&&  \!\!\!\!\! \!\!\!\!\!\VEV{\psi_1^d}  =\mathrm{e}^{-2 i \gamma k} \left(  v^d +\delta v_1^d \right) \; , \;\; 
\VEV{\psi_2^d}  =  v^d +\delta v_2^d \; , \;\;
\VEV{\chi_1^d}  =  \mathrm{e}^{-2 i \gamma k} \left(w^d +\delta w_1^d\right) \; , \;\;
\VEV{\chi_2^d}  = \mathrm{e}^{2 i \gamma k} \left(w^d + \delta w_2^d\right) \; , \;\; \\ 
&&\label{eq:VEVshifts}
  \!\!\!\!\! \!\!\!\!\!\VEV{\xi_1^d}  =  \mathrm{e}^{- 4 i \gamma k}\left(z^d+\delta z_1^d\right) \; , \;\; 
\VEV{\xi_2^d} =  \mathrm{e}^{4 i \gamma k}\left(z^d+\delta z_2^d\right) 
\;\;\; \mbox{and} \;\;\;
\VEV{\eta^d} =  \mathrm{e}^{-8 i \gamma k} \left( \frac{e_d}{f_d} \frac{v^d z^d}{w^d} + \delta \eta^d \right) 
\; ,
\end{eqnarray}
with $\VEV{\sigma} = x$ undetermined
and we confirm that also in the present setup the shifts of the VEVs are of the generic order $\epsilon^2 \, \Lambda$, thus relatively suppressed to the leading VEVs
by a factor $\epsilon$.
The VEVs of the driving fields are determined by the equations associated with the $F$-terms of the flavons and the latter vanish trivially, if the VEVs of all driving fields vanish.

\mathversion{bold}
\section{Comment on relevant $D_{14}$ subgroups}
\mathversion{normal}
\label{sec:D14sub}

As emphasized, we derive the Cabibbo angle $\theta_C$ in the quark sector through the 
breaking of $D_{14}$ to a particular type of $Z_2$ groups in the up and the down quark sectors. Also, the result
that $\theta_{13}^\nu$ and $\theta_{23}^\nu$ deviate by $\mathcal{O}(\lambda)$ from their
$\mu\tau$ symmetric values, $\theta_{13}^\nu=0$ and $\theta_{23}^\nu =\pi/4$, is related
to the different $Z_2$ subgroups governing the right-handed neutrino mass matrix and the Dirac neutrino mass matrix.
This type of $Z_{2}$ subgroup is generated by an element
of the form $\mathrm{B} \, \mathrm{A}^{k}$ for $k$ being an integer between 0 and 13 (for definition of the generators $\mathrm{A}$ and $\mathrm{B}$ see \Appref{app:grouptheory}). 
Apart from fields in the trivial representation $\MoreRep{1}{1}$, flavons transforming as $\MoreRep{1}{3}$ ($\MoreRep{1}{4}$) are allowed to have a non-vanishing
VEV for $k$ being even (odd), because $\mathrm{B} \mathrm{A}^k=1$ for $\MoreRep{1}{3}$,  $k$ even and for $\MoreRep{1}{4}$, $k$ odd, respectively.
In the case of two fields $\varphi_{1,2}$ which form a doublet
$\MoreRep{2}{j}$ a $Z_{2}$ group generated by $\mathrm{B} \, \mathrm{A}^{k}$ is preserved, if
\footnote{One can check that the subgroup preserved by VEVs of the form given in \Eqref{eq:VEVstructure}
cannot be larger than $Z_2$, if the index $\rm j$ of the representation $\MoreRep{2}{j}$ is odd, i.e. the representation is faithful. 
For an even index $\rm j$ the subgroup is a $D_2$ group generated by the two elements $\rm A^7$
and $\mathrm{B} \mathrm{A}^k$ with $k$ being an integer between $0$ and $6$.}
\begin{equation}
\label{eq:VEVstructure}
\left( \begin{array}{c} \langle \varphi_{1} \rangle \\
 \langle \varphi_{2} \rangle \end{array} \right) \propto \left(
\begin{array}{c} \mathrm{e}^{-2 \, i \, \gamma \, \mathrm{j} \, k}\\ 1
\end{array}
\right) \; .
\end{equation}
Using the fact that fields transforming as $\MoreRep{1}{3}$ can only preserve $Z_{2}$ subgroups generated 
by $\mathrm{B} \, \mathrm{A}^{k}$ with $k$ even and flavons in $\MoreRep{1}{4}$ only those with
$k$ odd, it is possible to ensure that the $Z_2$ subgroup conserved in the up quark is different from
the one in the down quark sector, needed for a non-trivial Cabibbo angle \cite{D14_quarks}.
For the neutrinos we require that the $Z_2$ symmetry relevant for the right-handed neutrino mass matrix
coincides with $Z_2^d$, preserved in the down quark sector, whereas the $Z_2$ symmetry relevant for the Dirac neutrino mass matrix
is $Z_2^u$, conserved in the up quark sector. Then, a deviation from $\mu\tau$ symmetric mixing
of order $\mathcal{O}(\lambda)$ is achieved. In particular, one can check that neutrino mixing is $\mu\tau$ symmetric, if one of the two matrices is at LO determined by contributions
which leave the whole group $D_{14}$ invariant, while the other matrix preserves either $Z_2^u$ or $Z_2^d$. \footnote{Also, in the (hypothetical) case 
in which only up-type or down-type flavons  dominantly generate the right-handed neutrino
mass matrix and, at the same time, the Dirac neutrino mass matrix, the mixing in the neutrino sector will be $\mu\tau$ symmetric.} 
In such a situation the deviations from $\theta_{13}^\nu=0$ and $\theta_{23}^\nu=\pi/4$ arise from subleading corrections which
are generically suppressed by the small expansion parameter $\epsilon \approx \lambda^2 \approx 0.04$ and as a consequence
lead to a too small value of the reactor mixing angle.
The solar mixing angle is in all cases not fixed, but depends on the parameters 
of the neutrino mass matrix in Eq. (\ref{eq:mlight_LO}).

In the charged lepton sector a different type of $Z_2$ subgroup of $D_{14}$ remains intact because 
the fields $\chi^e_{1,2}$ form an unfaithful representation of the group. We recall that for an 
unfaithful representation the number of group elements which is represented by the identity matrix is larger than one.
In the case of the representation $\MoreRep{2}{2}$ not only the trivial element, but also the element $\mathrm{A}^7$ is represented by the identity matrix, 
see \Eqref{eq:generators} with $\mathrm{j}=2$ (this holds for every even $\mathrm{j}$). As a consequence,
any non-trivial VEV of the fields $\chi^e_{1,2}$ preserves a $Z_2$ subgroup generated by the element $\mathrm{A}^7$. 
This is in contrast to what happens in the case of the $Z_2$ subgroups generated by $\mathrm{B} \mathrm{A}^k$, because
in the latter case the alignment as shown in \Eqref{eq:VEVstructure} is crucial. 
Apart from flavons in unfaithful two-dimensional representations also fields transforming
as $\MoreRep{1}{1}$ and $\MoreRep{1}{2}$ leave the $Z_2$ group generated by $\mathrm{A}^7$ invariant, because $\mathrm{A}^7=1$.
For details and a more general discussion of the subgroups of dihedral groups see \cite{dntheory}.

\section{Summary}
\label{sec:summary}

We have presented a model in the framework of the MSSM with the flavor symmetry $D_{14}$ which predicts the Cabibbo angle to 
be $|V_{us}| \approx \sin \gamma \approx \lambda \approx 0.22$ and the angles $\theta_{13}^\nu$ and $\theta_{23}^\nu$ to deviate
by $\sin\gamma \approx \lambda$ from $\mu\tau$ symmetric mixing, $\theta_{13}^\nu=0$ and $\theta_{23}^\nu=\pi/4$, in the neutrino sector at LO. These predictions arise from a particular
breaking of the group $D_{14}$, namely
the mismatch between the two different $Z_2$ subgroups $Z_2^u$ and $Z_2^d$, generated through elements
of the form $\mathrm{B} \mathrm{A}^k$ with $k$ being either an even ($Z^u_2$) or an odd ($Z_2^d$) integer.
In the quark sector, the symmetry $Z_2^u$ determines the up quark mass matrix, while $Z^d_2$ the down quark mass matrix.
In the neutrino sector, the right-handed neutrino mass matrix is governed by the symmetry $Z_2^d$ at LO, 
whereas the Dirac neutrino mass matrix is governed by $Z_2^u$ at LO instead.
 In the charged lepton sector another flavon is
responsible for the $D_{14}$ breaking whose vacuum leaves another type of $Z_2$ symmetry intact, generated by the element $\rm A^7$. 
The different symmetry breaking sectors are separated with the help of an additional $Z_7$ symmetry.
The contribution of the charged lepton sector to the lepton mixing angles is at maximum $\epsilon \approx \lambda^2$.
Thus, the latter deviate from $\mu\tau$ symmetric mixing by $\mathcal{O}(\lambda)$,
$\theta_{13}^l \approx \mathcal{O}(\lambda)$ and $\theta_{23}^l - \pi/4\approx \mathcal{O}(\lambda)$. Especially, the value of the reactor mixing angle
is well compatible with the recent experimental indications \cite{T2K,MINOS,DC,DayaBay,RENO} and the global fit results \cite{fogli,maltoni,schwetz}. The mixing angles $\theta_{13}^q$ and $\theta_{23}^q$ in the 
quark sector and the Jarlskog invariant can be correctly accommodated. The solar mixing angle is generically of order one in our model. 
Charged fermion mass hierarchies are correctly reproduced  without fine-tuning. Light neutrino masses are dominantly generated through the type I seesaw 
mechanism and can have either hierarchy.
We have studied in detail the effects of NLO operators and we have shown that they only slightly perturb the LO results.

\subsection*{Acknowledgments}

CH would like to thank Ferruccio Feruglio for useful discussions. DM acknowledges MIUR (Italy) for financial support under the 
program "Futuro in Ricerca 2010 (RBFR10O36O)". DM also acknowledges the Physics Department of the University of Padova for 
their hospitality during the preparation of this work.

\appendix

\mathversion{bold}
\section{Group theory of $D_{14}$}
\mathversion{normal}
\label{app:grouptheory}

We briefly review the basic features of the dihedral group $D_{14}$.
Its order is 28, and it has four one-dimensional irreducible representations which we denote as
$\MoreRep{1}{i}$, $\rm i=1,...,4$ and six two-dimensional ones called $\MoreRep{2}{j}$,
$\rm j=1,...,6$. All of them are real and the representations $\MoreRep{2}{j}$
with an odd index $\rm j$ are faithful. The group is generated by the two elements
$\rm A$ and $\rm B$ which fulfill the relations \cite{dngrouptheory}
\begin{equation}
\label{eq:genrelations}
\mathrm{A}^{14} =\mathbb{1} \;\;\; , \;\;\; \rm B^2=\mathbb{1} \;\;\; ,
\;\;\; \rm ABA=B \; . 
\end{equation}
The generators $\rm A$ and $\rm B$ of the one-dimensional representations read
\begin{eqnarray}\nonumber
\MoreRep{1}{1} &\;\;\; : \;\;\;& \rm A=1 \; , \;\; B=1 \;\;\; ,\\  \nonumber
\MoreRep{1}{2} &\;\;\; : \;\;\;& \rm A=1 \; , \;\; B=-1  \;\;\; ,\\  \nonumber
\MoreRep{1}{3} &\;\;\; : \;\;\;& \rm A=-1 \; , \;\; B=1 \;\;\; ,\\  
\MoreRep{1}{4} &\;\;\; : \;\;\;& \rm A=-1 \; , \;\; B=-1 \; .
\end{eqnarray}
For the representation $\MoreRep{2}{j}$ they are represented by two-by-two matrices of the form
\begin{equation}
\label{eq:generators}
\rm A =\left(\begin{array}{cc} 
                           \mathrm{e}^{2 \, i \, \gamma \, \mathrm{j}} & 0 \\
                            0 & \mathrm{e}^{-2 \, i \, \gamma \, \mathrm{j}} 
          \end{array}\right) \; , \; \rm B=\left(\begin{array}{cc} 
                                       0 & 1 \\
                                       1 & 0 
                  \end{array}\right) \; .
\end{equation}
Note that we have chosen $\rm A$ to be complex, although all representations of $D_{14}$ are real.
Due to this, the combination $(a_{2}^{\star}, a_{1}^{\star})^{T}$ transforms as $\MoreRep{2}{j}$ for $(a_{1},a_{2})^{T}$ forming the doublet $\MoreRep{2}{j}$.

We list the explicit form of the Kronecker products as well as the Clebsch Gordan coefficients.
More general results for dihedral groups with an arbitrary index $n$ can be found in 
\cite{kronprods,dntheory}.

The products $\MoreRep{1}{i} \times \MoreRep{1}{j}$ are
\begin{equation}\nonumber
\MoreRep{1}{i} \times \MoreRep{1}{i}= \MoreRep{1}{1} \; , \;\; 
\MoreRep{1}{1} \times \MoreRep{1}{i}= \MoreRep{1}{i} \;\; \mbox{for} \;\; \rm i=1,...,4 
\; , \;\;
\MoreRep{1}{2} \times \MoreRep{1}{3}= \MoreRep{1}{4} \; , \;\;
\MoreRep{1}{2} \times \MoreRep{1}{4}= \MoreRep{1}{3} \;\; \mbox{and} \;\;
\MoreRep{1}{3} \times \MoreRep{1}{4}= \MoreRep{1}{2} \; .
\end{equation}
For $\MoreRep{1}{i} \times \MoreRep{2}{j}$ we find
\begin{equation}\nonumber
\MoreRep{1}{1,2} \times \MoreRep{2}{j}= \MoreRep{2}{j} \;\;\; \mbox{and} \;\;\;
\MoreRep{1}{3,4} \times \MoreRep{2}{j}= \MoreRep{2}{7-j} \;\;\; \mbox{for all} \;\;\; \rm j \; .
\end{equation}
The products of $\MoreRep{2}{i} \times \MoreRep{2}{i}$ decompose into
\begin{equation}\nonumber
\left[ \MoreRep{2}{i} \times \MoreRep{2}{i} \right] = \MoreRep{1}{1} + \MoreRep{2}{j}
\;\;\; \mbox{and} \;\;\;
\left\{ \MoreRep{2}{i} \times \MoreRep{2}{i} \right\} = \MoreRep{1}{2} \, ,
\end{equation}
where the index $\rm j$ equals $\rm j=2i$ for $\rm i \leq 3$ and 
$\mathrm{j}= 14 - 2\rm i$ holds for $\rm i \geq 4$. $\left[ \nu \times \nu \right]$ denotes 
the symmetric part of the product $\nu \times \nu$, while $\left\{ \nu \times \nu \right\}$ is the anti-symmetric 
one. For the mixed products $\MoreRep{2}{i} \times \MoreRep{2}{j}$ with $\rm i \neq j$
two structures are possible. For $\rm i+j\neq 7$ it is
\begin{equation}\nonumber
\MoreRep{2}{i} \times \MoreRep{2}{j} = \MoreRep{2}{k} + \MoreRep{2}{l}
\end{equation}
with $\rm k=|i-j|$ and $\rm l$ being $\rm i+j$ for $\rm i+j \leq 6$ and $14 - (\rm i+j)$ for $\rm i+j \geq 8$.
For $\rm i+j=7$ we find instead
\begin{equation}\nonumber
\MoreRep{2}{i} \times \MoreRep{2}{j} = \MoreRep{1}{3} + \MoreRep{1}{4} + \MoreRep{2}{k} \, ,
\end{equation}
where $\rm k$ is again $\rm |i-j|$.

The Clebsch Gordan coefficients for a product of a one-dimensional representation, $s_i \sim \MoreRep{1}{i}$, with a two-dimensional
one, $(a_1,a_2)^{T} \sim \MoreRep{2}{j}$, are 
\begin{equation}\nonumber
\left( \begin{array}{c} s_1 a_1 \\ s_1 a_2
\end{array} \right) \sim \MoreRep{2}{j} \;\; , \;\;\;
\left( \begin{array}{c} s_2 a_1 \\ -s_2 a_2
\end{array} \right) \sim \MoreRep{2}{j} \;\; , \;\;\;
\left( \begin{array}{c} s_3 a_2 \\ s_3 a_1
\end{array} \right) \sim \MoreRep{2}{7-j} \;\;\; \mbox{and} \;\;\;
\left( \begin{array}{c} s_4 a_2 \\ -s_4 a_1
\end{array} \right) \sim \MoreRep{2}{7-j} \;\; .
\end{equation}
The Clebsch Gordan coefficients of the product of $(a_{1},a_{2})^{T}$, $(b_{1},b_{2})^{T}$ 
$\sim \MoreRep{2}{i}$ read
\begin{equation}\nonumber
a_{1} b_{2} + a_{2} b_{1} \sim \MoreRep{1}{1} \; , \;\;
 a_{1} b_{2} - a_{2} b_{1} \sim \MoreRep{1}{2} \; , 
\;\;\;  \left( \begin{array}{c}
	a_{1} b_{1}\\
	a_{2} b_{2}
	\end{array}
	\right) \sim \MoreRep{2}{j} \;\;\; \mbox{or} \;\;\;
 \left( \begin{array}{c}
	a_{2} b_{2}\\
	a_{1} b_{1}
	\end{array}
	\right) \sim \MoreRep{2}{j}
\end{equation}
depending on whether $\rm j= 2 i$ as it is for $\rm i \leq 3$
or $\mathrm{j}= 14 - 2 \rm i$ which holds if $\rm i \geq 4$.
For the two doublets $(a_{1}, a_{2})^{T} \sim \MoreRep{2}{i}$ and
$(b_{1}, b_{2})^{T} \sim \MoreRep{2}{j}$ we find for 
$\rm i + j \neq 7$
\begin{eqnarray}\nonumber
&& \left( \begin{array}{c}
	a_{1} b_{2}\\
	a_{2} b_{1}
	\end{array}
	\right) \sim \MoreRep{2}{k} \;\;\; (\rm k=i-j) \;\;\; \mbox{or} \;\;\;
 \left( \begin{array}{c}
	a_{2} b_{1}\\
	a_{1} b_{2}
	\end{array}
	\right) \sim \MoreRep{2}{k} \;\;\; (\rm k=j-i)
\\ \nonumber
&& \left( \begin{array}{c}
	a_{1} b_{1}\\
	a_{2} b_{2}
	\end{array}
	\right) \sim \MoreRep{2}{l} \;\;\;  (\rm l=i+j) \;\;\;\;\; \mbox{or} \;\;\;
 \left( \begin{array}{c}
	a_{2} b_{2}\\
	a_{1} b_{1}
	\end{array}
	\right) \sim \MoreRep{2}{l} \;\;\; (\rm l=14 -(i+j)) \, .
\end{eqnarray}
If $\rm i+j=7$ holds, the co-variants are
\begin{equation}\nonumber
a_{1} b_{1} + a_{2} b_{2} \sim \MoreRep{1}{3} \; , \;\;
 a_{1} b_{1} - a_{2} b_{2} \sim \MoreRep{1}{4} \; , 
\;\;\; \left( \begin{array}{c}
	a_{1} b_{2}\\
	a_{2} b_{1}
	\end{array}
	\right) \sim \MoreRep{2}{k} \;\;\; \mbox{or} \;\;\;
 \left( \begin{array}{c}
	a_{2} b_{1}\\
	a_{1} b_{2}
	\end{array}
	\right) \sim \MoreRep{2}{k} \; .
\end{equation}
Again, the first case is relevant for $\rm k=i-j$, while the second form for
$\rm k=j-i$.



\end{document}